## Journal of Geophysical Research: Space Physics

# Characteristics of the flank magnetopause: Cluster observations

**S. Haaland[1,2], J. Reistad[2], P. Tenfjord[2], J. Gjerloev[2,3], L. Maes[4], J. DeKeyser[4], R. Maggiolo[4], C. Anekallu[5], and N. Dorville[6]**

[1]Birkeland Centre for Space Science, University of Bergen, Bergen, Norway, [2]Max-Planck Institute for Solar Systems Research, Göttingen, Germany, [3]Applied Physics Laboratory, Johns Hopkins University, Baltimore, Maryland, USA, [4]Belgian Institute of Space Aeronomy, Brussels, Belgium, [5]Mullard Space Science Laboratory, Dorking, UK, [6]LPP, Ecole Polytechnique, Université Paris-Sud, Palaiseau, France

**Abstract** The magnetopause is a current sheet forming the boundary between the geomagnetic field on one side and the shocked solar wind on the other side. This paper discusses properties of the low-latitude dawn and dusk flanks of the magnetopause. The reported results are based on a large number of measurements obtained by the Cluster satellites during magnetopause traversals. Using a combination of single-spacecraft and multispacecraft techniques, we calculated macroscopic features such as thickness, location, and motion of the magnetopause. The results show that the typical flank magnetopause is significantly thicker than the dayside magnetopause and also possesses a pronounced and persistent dawn-dusk asymmetry. Thicknesses vary from 150 to 5000 km, with an median thickness of around 1400 km at dawn and around 1150 km at dusk. Current densities are on average higher on dusk, suggesting that the total current at dawn and dusk are similar. Solar wind conditions and the interplanetary magnetic field cannot fully explain the observed dawn-dusk asymmetry. For a number of crossings we were also able to derive detailed current density profiles. The profiles show that the magnetopause often consists of two or more adjacent current sheets, each current sheet typically several ion gyroradii thick and often with different current direction. This demonstrates that the flank magnetopause has a structure that is more complex than the thin, one-dimensional current sheet described by a Chapman-Ferraro layer.

## 1. Introduction

The terrestrial magnetopause marks the boundary between the geomagnetic field on one side and the shocked solar wind with its embedded interplanetary magnetic field (IMF) on the other side. It is thus a key region for transfer of mass, momentum, and energy from the solar wind into the magnetosphere. Due to this vital role, the magnetopause has been the target of in situ measurements since the beginning of space age. One of the first unambiguous observations of the magnetopause current layer was made by the Explorer 12 spacecraft in 1963 [*Cahill and Amazeen*, 1963]. Subsequent observations by the OGO [e.g., *Aubry et al.*, 1970] and ISEE spacecraft [e.g., *Russell and Elphic*, 1978; *Berchem and Russell*, 1982] provided additional knowledge about orientation, motion, and thickness of the dayside magnetopause current sheet.

To the first order, the magnetopause has often been considered as a thin current sheet (sometimes referred to as a Chapman-Ferraro current, after *Chapman and Ferraro* [1930]) set up by the opposite motion of electrons and ions as they encounter the sharp gradient at the boundary. From Ampères's law, it follows that such a current sheet will be observed as an abrupt change in the magnetic field magnitude and a rotation of the magnetic field vector by a spacecraft traversing it.

Since the magnetopause position is dictated by the total pressure balance between the plasmas on either side, and given the presence of pressure variations in the solar wind at all timescales, the magnetopause is constantly moving back and forth. Consequently, spacecraft typically observe multiple encounters of the magnetopause current sheet. The European Space Agency's (ESA) Cluster multispacecraft mission has been instrumental in disentangling this motion from the spatial structure of the boundary [e.g., *De Keyser et al.*, 2005; *Paschmann et al.*, 2005].

Investigations have also revealed that the magnetopause is home to a number of processes that can cause transport of plasma from the solar wind into the magnetosphere. Reconnection [e.g., *Paschmann et al.*, 1979; *Sonnerup et al.*, 1981] is presumably the most prominent process, but also diffusion [e.g., *Tsurutani* *and Thorne*, 1982; *Treumann*, 1999; *Treumann and Sckopke*, 1999], impulsive penetration [e.g., *Lemaire*, 1977; *Lundin et al.*, 2003], and nonlinear Kelvin-Helmholtz (KH) waves [e.g., *Southwood*, 1979; *Ogilvie and Fitzenreiter*, 1989; *Hasegawa et al.*, 2004a] have been suggested to enable transport across the magnetopause. Reconnection usually requires large magnetic shear, typically between the IMF and the geomagnetic field. But *Hasegawa et al.* [2009] demonstrated that KH waves can initiate local reconnection when vortices roll over and create local regions with large shear.

With the availability of better observations and new techniques, it has also been possible to study small-scale and internal structures of the magnetopause in more detail. Such studies have revealed that magnetopause often exhibit internal structures such as magnetic islands [e.g., *Hasegawa et al.*, 2004b; *Teh et al.*, 2010] and layered current sheets [e.g., *Haaland et al.*, 2004].

Much of the attention has been focused on the dayside magnetopause, since processes (in particular reconnection) in this region have the largest impact on the dynamics of the magnetosphere. The Dungey cycle [*Dungey*, 1961]—the large-scale circulation of plasma in the magnetosphere and the magnetically connected ionosphere—is a direct consequence of the interaction between the IMF and the geomagnetic field at the dayside magnetopause.

The flanks of the magnetopause and possible asymmetries between the dawnside and duskside have received less attention, partly because interactions along the flanks have less direct effect of geoactivity, and partly because there are less observations from this region.

It has been known for some time that properties and processes in the adjacent magnetosheath possess dawn-dusk asymmetries [e.g., *Walters*, 1964] (see also a recent review by *Walsh et al.* [2013]). Some of these asymmetries follow from the interaction between the bow shock with the predominantly Parker spiral like IMF orientation [*Parker*, 1958; *McCracken*, 1962]. Due to the velocity shear between the magnetopause and the magnetosheath at the flanks, Kelvin-Helmholtz waves are often observed along the interface. *Taylor et al.* [2012] noted that such KH waves are more frequent on the dusk flank magnetopause.

Conditions inside the magnetosphere may also play a role. Return feed of plasma from the plasmasphere in the form of a plasmaspheric plume can also contribute to a dawn-dusk asymmetry at the magnetopause [e.g., *Borovsky and Denton*, 2006; *Walsh et al.*, 2014].

Furthermore, as the magnetopause and its boundary layer are magnetically coupled to the ionosphere, any dawn-dusk asymmetries in ionospheric conductivity and ionospheric electric potential near the field line foot points and/or differences in the field-aligned current-voltage relations may also lead to magnetopause asymmetries [e.g., *Atkinson and Hutchison*, 1978; *Tanaka*, 2001].

Apart from asymmetries in boundary conditions, the physics of the magnetopause current sheet itself implies an asymmetric response dictated by the orientation of the flow shear relative to the magnetic field. The fundamental reason for this asymmetry is the different response of ions and electrons in the Chapman-Ferraro current layer to the additional electric field that stems from the cross-field flow and that has a different sense in the dawn and dusk configurations. This effect has implications for boundary thickness, for the sense of magnetic field rotation, and it actually even sets limits on the regions where equilibrium tangential discontinuity equilibria for the magnetopause are feasible [*De Keyser and Roth*, 1998a, 1998b].

In the present study, which is an extension and follow up of *Haaland and Gjerloev* [2013], we make use of more than 10 years of observations from the Cluster satellites to investigate macroscopic features of the dawn and dusk magnetopause flanks. In particular, we examine to what degree external drivers such as the solar wind and IMF exert control over the flank magnetopause. Having a large number of observations from both flanks of the magnetopause, we also present experimental evidence for a persistent dawn-dusk asymmetry in many of the parameters. In this paper we focus on observations of magnetopause properties, with emphasis on thickness, normal velocity, and current density. A future companion paper (L. Maes et al., in preparation) will discuss possible mechanisms for the asymmetry, and another companion paper in progress (C. Anekallu, in preparation) will focus on magnetopause orientation and fit to models.

The paper is organized as follows. In section 2 we provide a brief description of the data set used and describe how the magnetopause traversals were identified. Section 3 presents examples of magnetopause observations and gives an overview of the methodology to determine key magnetopause parameters.

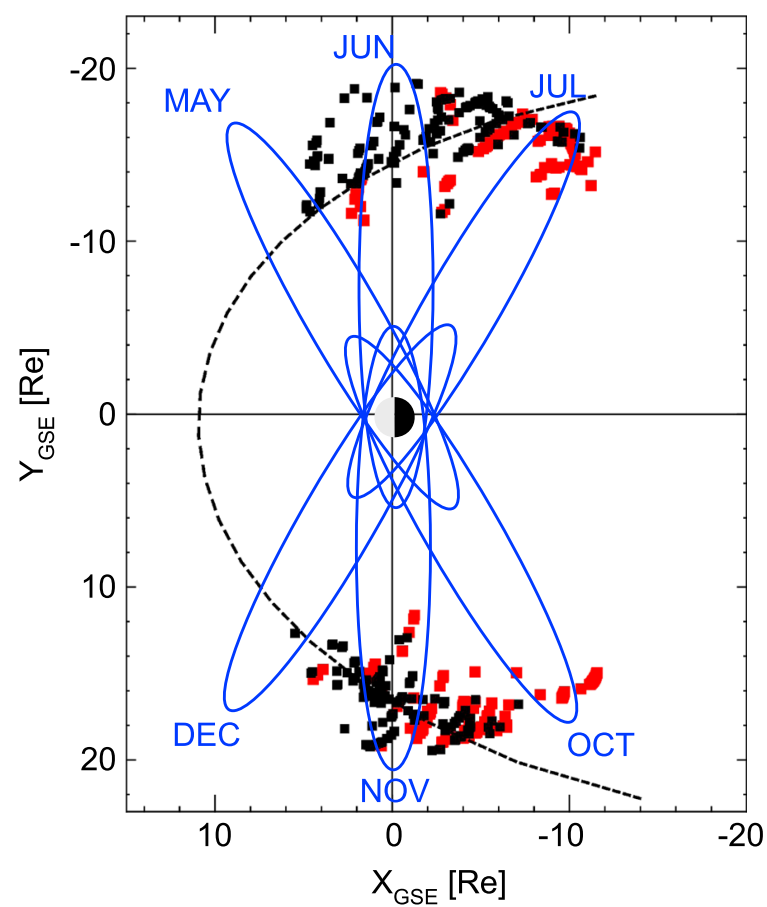

**Figure 1.** Cluster orbits during the years 2001–2010 used in this study. Cluster traversed the dawn magnetopause during May to July and the dusk flank during October to December. Observations from these periods are used to calculate macroscopic magnetopause parameters for the two regions. The dashed curve indicates a 4° aberrated magnetopause [*Fairfield*, 1971], parameterized with average solar wind parameters from our data set. Black and red dots indicate actual Cluster C3 magnetopause crossings fulfilling certain quality criteria and will be discussed further in section 4.

In section 4 we present statistical results of magnetopause parameters and their correlation with solar wind parameters. Finally, in section 5 we summarize the findings of the study.

## 2. Data

The data basis for the present study consists of almost 5800 magnetopause traversals by one or more of the four Cluster satellites during the years 2001–2010. In the science community, the four satellites are usually referred to as C1, C2, C3, and C4, and we will use this notation in this paper. Investigation of boundaries in space was one of the prime objectives of the Cluster mission. During the first years of operation (approximately 2001–2006), the spacecraft were flying in a tetrahedron-like formation with varying separation distance. After 2006, the spacecraft configuration changed so that one pair of spacecraft often lead the others with a large separation distance between the pairs.

All four Cluster spacecraft have the same instrumentation, but not all instruments work on all spacecraft. In this study, we have mainly used data from the flux gate magnetometer [see *Balogh et al.*, 2001] and plasma data from the Cluster Ion Spectrometry (CIS) experiment [see *Rème et al.*, 2001]. For more information about Cluster, including mission objectives, instrumentation, and operations, we refer to *Escoubet et al.* [2001]. A good overview of Clusters contribution to the study of the Earth's outer magnetospheric boundary layers can be found in *Paschmann et al.* [2005].

Of particular interest for our study are the traversals near the low latitude (±45° of the geocentric solar ecliptic (GSE) plane) dawn and dusk magnetopause. Cluster traverses the dawn magnetopause during the period from mid-May to mid-July, and the dusk flank during the months October to early December as illustrated in Figure 1. Due to aberration caused by Earth's orbital motion (approximately 30 km s$^{-1}$) around the Sun, the magnetopause at dawn is somewhat closer to the Cluster apogee (around 19 $Re$) than its dusk counterpart. We therefore have more observations at dawn. Also, since Cluster's line of apsides moves southward with time, flank magnetopause crossings in the later years take place at high (Southern Hemisphere) latitudes, and the ±45° criterion is often not fulfilled. For the same reason, there are also more magnetopause (MP) crossings in the Southern Hemisphere than in the Northern Hemisphere. With equinoxes in late March and late September, there should be no tilt angle bias in this data set.

The selection of data for this study largely follows the same scheme as that of *Haaland and Gjerloev* [2013]. For convenience, we provide a brief description below.

### 2.1. Identifying Cluster Magnetopause Traversals

Magnetopause crossings were selected in a multistep procedure. First, the Cluster quick look pages (at time of writing, these were accessed trough the URL http://www.cluster.rl.ac.uk/csdsweb-cgi/csdsweb_pick) were consulted to identify approximate times for magnetopause traversals. These www-pages allow users to plot low-time resolution key parameters such as plasma moments, fields, and spacecraft position. Magnetopause traversals were visually identified from abrupt changes in field or plasma parameters which could indicate a transition between the fairly rigid magnetic field and low-plasma density inside the magnetosphere to a more turbulent magnetic field and higher-plasma densities in the magnetosheath.

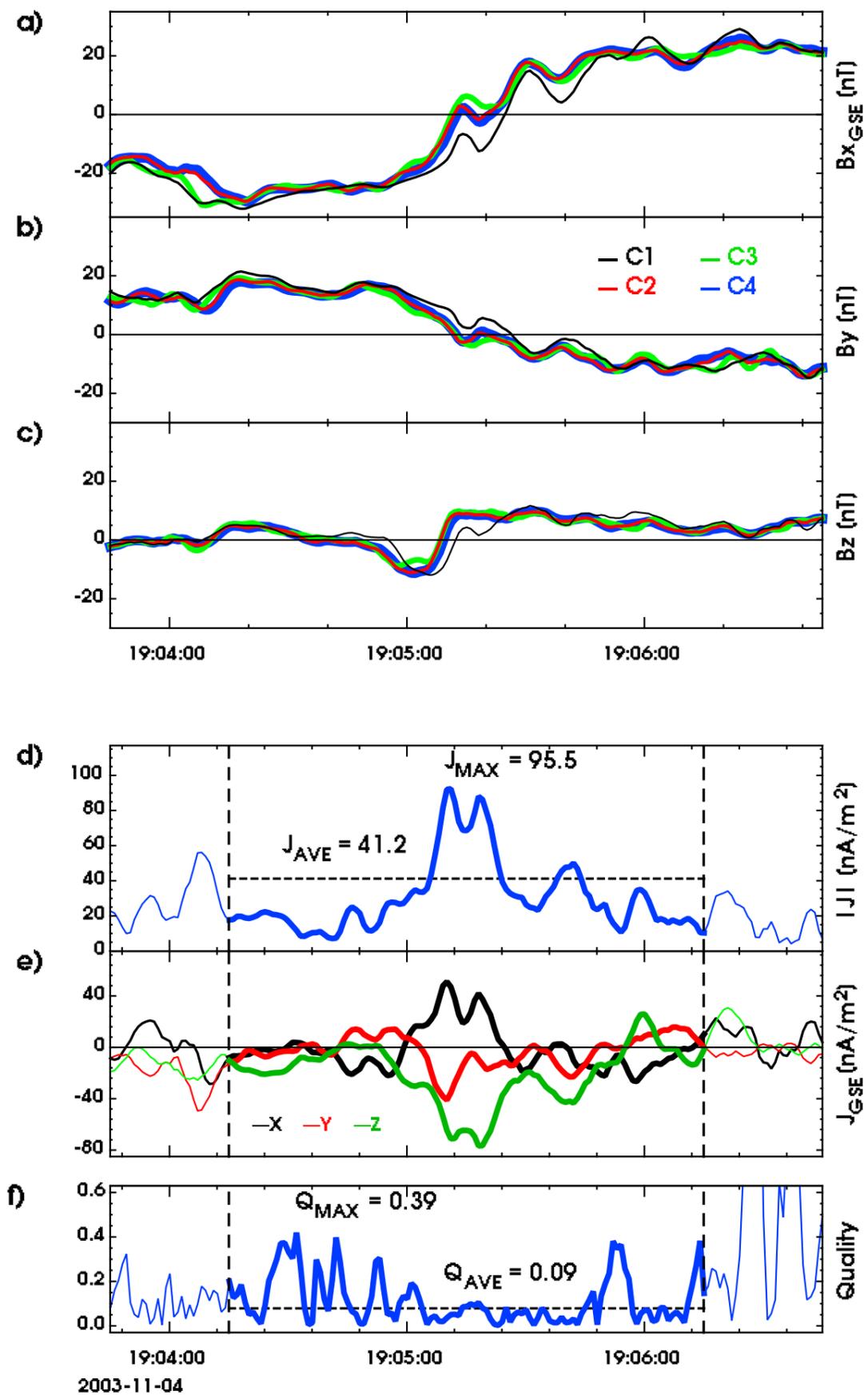

**Figure 2.** Magnetopause current sheet profile during a duskside magnetopause crossing on 4 November 2003. (a–c) High-resolution GSE components of the magnetic field; standard Cluster color coding has been used and is indicated in Figure 2b. (d) The calculated current density, which exhibits a pronounced two-peak structure in this case. (e) The individual components of the current. (f) The quality criteria, $Q = |\nabla \cdot \vec{B}|/|\nabla \times \vec{B}|$. Bold lines in Figures 2d–2f mark the time interval used to calculate average quantities used in statistics (see section 4).

Times of events where the spacecraft position, field and plasma parameters suggested a crossing of the magnetopause were noted and spin resolution data for a time interval around the suspected magnetopause traversal were downloaded. For each identified time period, we plotted the magnetic field, plasma density, and plasma flow (when available). From these overview plots, we then tried to identify the magnetopause traversal times to within a minute accuracy for further processing. To obtain a more precise timing of the magnetopause crossings, and also for calculation of current density, we plotted high-resolution (5 vec/s) magnetic field data downloaded from the Cluster Active Archive [*Laakso et al.*, 2010].

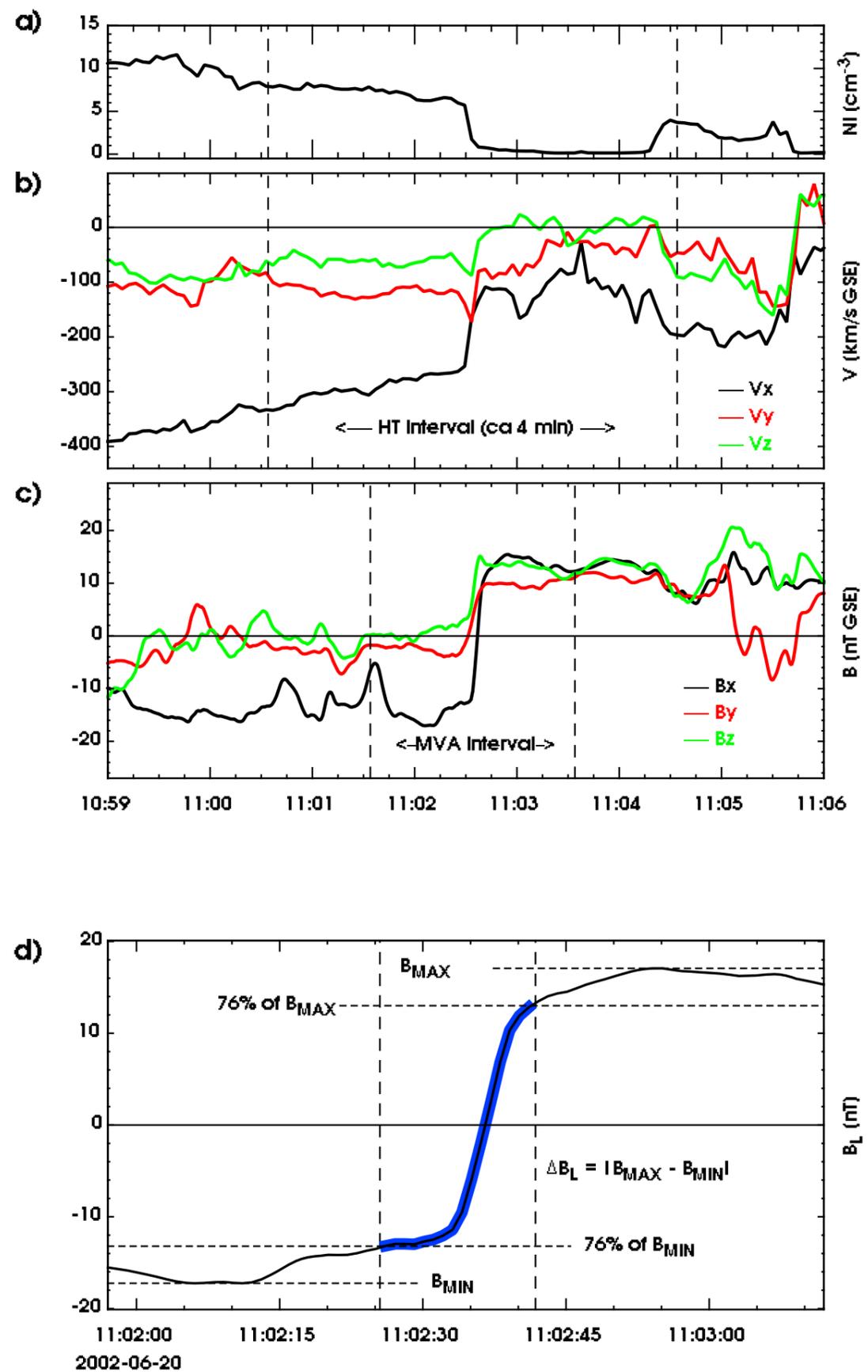

**Figure 3.** Example of a dawn magnetopause crossing on 20 June 2002 with some key elements used to calculate thickness and current from single-spacecraft methods. (a) Ion density. (b) Ion velocity used to calculate magnetopause motion. (c) Magnetic field components used to calculate the magnetopause normal. (d) Zoomed in interval with $B_L$ (maximum variance) component of the magnetic field and duration ($\Delta T$ = time it takes to perform 76% of the field rotation—blue part of line) and minimum and maximum values indicated. Vertical dashed lines indicate the time intervals used to calculate velocity and orientations, respectively. Adapted from *Haaland and Gjerloev* [2013].

Due to the oscillatory motion of the magnetopause, Cluster typically observes a number of inbound and outbound passages trough the magnetopause current layer during each orbit. At the end, we had visually identified around 5800 magnetopause traversals during the years 2001 to 2010, where one, two, or three, or all four Cluster spacecraft crossed the magnetopause.

**Table 1.** Filter Criteria Used to Exclude Records Where Magnetopause Parameters Could Not be Accurately Determined[a]

| Quality Criteria | Method | Allowed Range | Remarks |
|---|---|---|---|
| GSE latitude | Both | $\leq 45°$ | Avoid high-latitude or cusp crossings |
| Current density | Both | 0 to 200 nA m$^{-2}$ | Remove records with unrealistic magnetopause current density |
| Year | Curlometer | 2001–2006 | Spacecraft configuration not suitable for curlometer calculations after 2006. |
| $Q_{AVE}$ | Curlometer | $\leq 0.25$ | Curlometer quality estimate—see section 3.1 |
| $Q_{MAX}$ | Curlometer | $\leq 1.00$ | $\nabla \cdot \vec{B}$ should never exceed $\nabla \times \vec{B}$—see equation (2) |
| MP thickness | Single SC | 150 to 5000 km | Remove records with unrealistic magnetopause thickness |
| MVA eigenvalue ratio | Single SC | $\geq 10$ | Ensure well-determined magnetopause orientation |
| $\Delta B$ | Single SC | $\geq 10$ nT | Make sure jump in B field is 10 nT or more |
| HTcc | Single SC | $\geq 0.85$ | Ensure well-defined magnetopause velocity and thickness |

[a]With exception of an adjusted deHoffmann-Teller correlation criteria, these values are adapted from *Haaland and Gjerloev* [2013].

The number may seem large, but note that a concise definition of "magnetopause" in terms of universally agreed thresholds in field and plasma changes does not really exist. Furthermore, any study based on visual inspection is bound to be biased in some way. Even automated routines such as those used by e.g., *Ivchenko et al.* [2000] or *Case and Wild* [2013] require that the user defines the detection criteria. An observation classified as a magnetopause crossing by one observer will not necessarily be classified as such by another observer. Our data set should therefore *not* be considered as a *complete* catalog of Cluster magnetopause crossings for the given period.

Magnetic field rotations and strong plasma gradients can also occur in adjacent boundary layers and in the magnetosheath. For example, *Retinò et al.* [2007] reported observations of reconnection in thin current sheets embedded in the magnetosheath. During the analysis, we tried to minimize the effects of such issues by using subsets of the full data set where certain thresholds in the observed plasma and field changes were satisfied (see section 4). We also required that derived quantities such as eigenvalues, correlation coefficients, and quality factors were within certain limits. Depending on which macroscopic parameter we studied, the number of crossings used for a specific characterization therefore varies.

## 3. Methodology and Examples

One great advantage of the Cluster mission compared to earlier missions is the possibility to use data from all four spacecraft to resolve boundary layers in space. Although breakdown of some of the underlying assumptions (planarity, constancy in motion and linear variations) for four-spacecraft methods frequently occurs, it is still often possible to combine data from two or more spacecraft in the analysis.

The techniques to determine magnetopause parameters consist of a combination of multispacecraft and single-spacecraft methods. The curlometer technique [e.g., *Dunlop et al.*, 2002] has been used to obtain current densities wherever possible. Orientation, speed, and discontinuity classification are derived from single- or dual-spacecraft field and plasma measurements [e.g., *Sonnerup and Scheible*, 1998; *Sonnerup et al.*, 2008]. All these techniques have been widely applied for analysis of data from both Cluster and other missions, and are well-documented and extensively tested and benchmarked in a number of studies. We recommend the book by *Paschmann and Daly* [1998] for details about the methods and the successor [*Paschmann and Daly*, 2008] for extensive lists of papers where the various methods have been applied.

Below, we briefly explain the procedure used to determine magnetopause parameters using two examples from our data base.

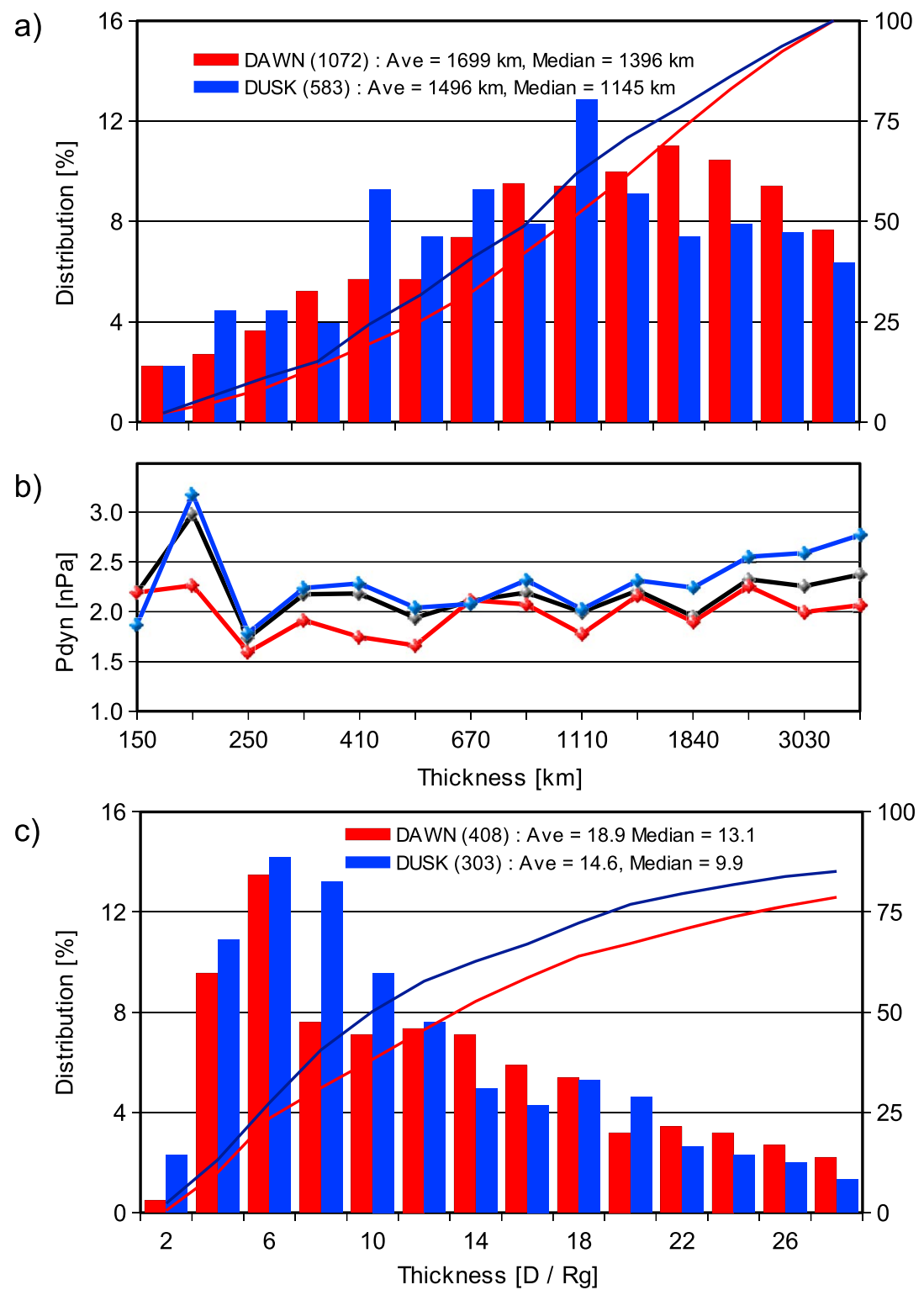

**Figure 4.** Magnetopause thickness characteristics. Blue color indicate dusk crossings, red color indicate dawn crossings. Note that we use (a and b) 14 logarithmically spaced bins and (c) 14 linearly spaced bins. Figure 4a are normalized histograms of magnetopause thickness in units of kilometer. Lines show corresponding cumulated values. Number of events, mean, and median values for each flank are indicated. Solid lines indicate accumulated values and attached to the secondary vertical axis at right. Figure 4b is magnetopause thickness versus solar wind dynamic pressure. We here also plot the combined (dawn+dusk) correlation as a black line. No significant dependence between magnetopause thickness and solar wind dynamic pressure is apparent. Figure 4c is the same as Figure 4a but with the thickness in terms of ion gyroradii.

### 3.1. Current Density

Provided that the spacecraft configuration is properly formed (see e.g., *Robert et al.* [1998] and *Chanteur and Harvey* [1998], for a discussion about this) and the spacecraft separation is small compared to the structure observed, differences in measurements between the spacecraft can be used to estimate linear gradients in 1, 2, or 3 dimensions. The curlometer method makes use of this to calculate the electric current density from Amperes law:

$$\mu_0 \vec{J} = \nabla \times \vec{B} \tag{1}$$

The same $\nabla$ operator can also be used to calculate the divergence of $\vec{B}$. This should ideally be zero, so the ratio

$$Q = \frac{|\nabla \cdot \vec{B}|}{|\nabla \times \vec{B}|} \qquad (2)$$

provides an estimate of the quality of the current determination. Small values of $Q$ are desirable; values of $Q$ close to, or larger than unity indicate a nontrustworthy current estimate. However, the divergence is also calculated from gradients and thus restricted by the above assumptions. The value of $Q$ can therefore not be used to derive exact error estimates of the current determination.

As part of the processing scheme, we calculated curlometer results for all crossings in the data set, but for a large number of crossings the spacecraft configuration was unsuitable. In particular, after 2006, the space-craft configuration was changed so that one or two of the Cluster spacecraft trailed the others with a large separation distance. Although variants of the curlometer for 2 or 3 spacecraft exist [e.g., *Vogt et al.*, 2009, 2013], stronger assumptions or additional models are required. The present data set therefore contains curlometer results for the years 2001–2006, i.e., the same time period as used in *Haaland and Gjerloev* [2013].

The curlometer method allows us to study the magnetopause current sheet profile in detail. Figure 2 shows an example of a layered current sheet profile observed at the dusk flank on 4 November 2003. Cluster was initially located inside the magnetopause, and entered the magnetopause current layer around 19:04:55 UT, and exited into the magnetosheath approximately 40 s later. The current profile shown in Figure 2d reveals a double-peaked structure with current density reaching 95 nA m$^{-2}$ near the magnetospheric side of the plot. The quality criteria, $Q = |\nabla \cdot \vec{B}|/|\nabla \times \vec{B}|$, shown in Figure 2f is reasonable during the whole crossing.

A closer look at the individual components of the current, shown in Figure 2e, reveals that the initial peak has a significant $y$ component (red line), not present in the second peak. We interpret this as the signature of two adjacent current sheets, with different current direction. This layered profile is by no means unique. Many of the events in our collection possessed similar signatures. The example illustrated in Figure 5 of *Haaland and Gjerloev* [2013] and the dayside crossing in *Haaland et al.* [2004] both indicate similar, layered current sheets.

The central current peak (i.e., the part of the blue line above the dashed average in Figure 2d) has a duration of approximately 20 s. Single-spacecraft methods (described in the next section) give an average magnetopause velocity of around 32 km s$^{-1}$. The thickness of the central current sheet is thus of the order of 600 km, corresponding to 8 ion gyroradii. Even the small individual current peaks indicate current layers several ion gyroradii thick.

### 3.2. Orientation, Motion, and Thickness

Magnetopause orientation, velocity, and thickness are obtained from single-spacecraft methods. Figure 3 shows an example of a dawnside magnetopause crossing observed by Cluster C3 on 20 June 2002. The spacecraft is initially in the magnetosheath, characterized by a high-plasma density (around 10 cm$^{-3}$—see Figure 3a), bulk flow speed of several hundreds km s$^{-1}$ (Figure 3b), and a fairly turbulent magnetic field (Figure 3c). Starting around 11:02:25, a rapid rotation of the magnetic field, combined with a drop in plasma density and velocity, is observed as the spacecraft enters the magnetopause current sheet. The duration of the field rotation is only about 15 s. We will use this event to illustrate how orientation, motion, and thickness of the magnetopause can be derived.

The magnetopause normal, $\vec{n}$, is calculated from a constrained minimum variance analysis (MVABC) [see *Sonnerup and Scheible*, 1998] of the magnetic field measurement taken across the magnetopause. The large number of crossings does not allow for any individual treatment of each crossing, so we have used a fixed 2 min interval for all crossings. MVABC is usually more stable and less sensitive to the time interval than standard minimum variance, though [*Sonnerup et al.*, 2008]. The underlying hypothesis is that the mag-netopause is a tangential discontinuity or, if it is a rotational discontinuity, that the normal component is relatively small. This hypothesis that may not necessarily be true for each individual case but is certainly valid in an average sense.

Constraining the variance analysis this way implies that the normal magnetic field vanishes. This provides us with a set of two nonzero eigenvalues which are used as a quality criterion. A large eigenvalue ratio ($\lambda_{max}/\lambda_{intermediate}$) is indicative of a well-determined boundary orientation. Note that the eigenvalue ratio for this constrained variance is not directly comparable to the $(\lambda_{intermediate}/\lambda_{min})$ ratio often used as a quality criterion for the standard, unconstrained minimum variance analysis.

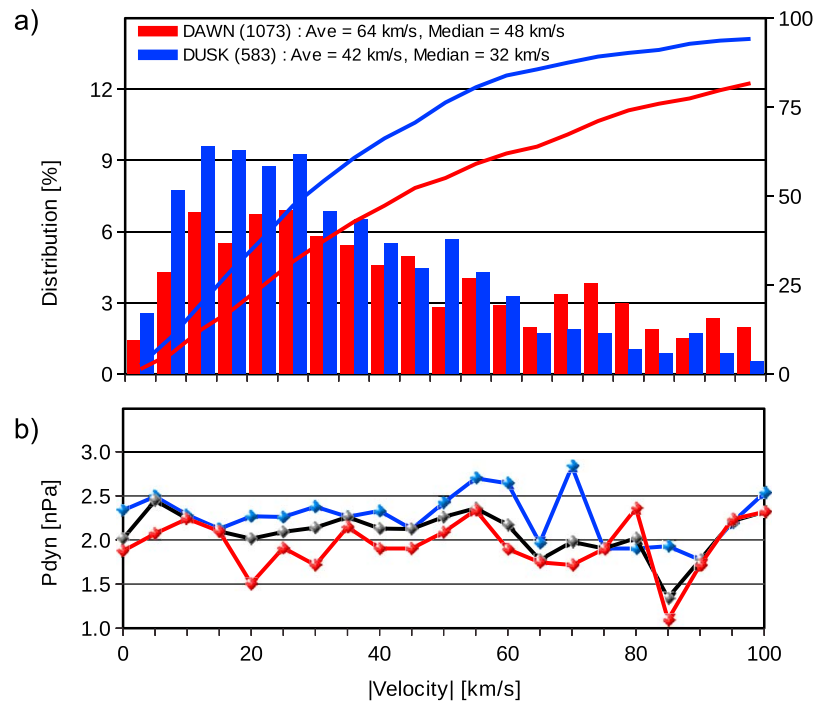

**Figure 5.** (a) Normalized histograms of magnetopause velocity. Lines show the corresponding cumulated value, and attached to the secondary vertical axis at right. Number of events, mean, and median values are indicated. (b) Magnetopause normal velocity versus solar wind dynamic pressure. The black line shows the corresponding dependence for the combined (dawn+dusk) crossings. As for the thickness, we find little or no correlation between magnetopause velocity and the solar wind dynamic pressure.

As magnetopause velocity, we use the deHoffmann-Teller frame velocity, $\vec{V}_{HT}$ [de Hoffmann and Teller, 1950; Paschmann and Sonnerup, 2008], calculated from ion moments and spin resolution magnetic field from Cluster C1 and C3 for a 4 min interval around the crossing. The frame velocity is thereafter projected along the boundary normal $\vec{n}$ to find the normal velocity $V_n = \vec{V}_{HT} \cdot \vec{n}$. Calculating $\vec{V}_{HT}$ consists of finding a frame in which the motional electric field, $\vec{E}' = \vec{E} + \vec{V}_{HT} \times \vec{B}$, is minimized. The correlation, HTcc, between the measured electric field $\vec{E}_c = -\vec{V} \times \vec{B}$ and $\vec{E}_{HT} = -\vec{V}_{HT} \times \vec{B}$ describes how well the frame is determined and thus provides us with a quality estimate of the magnetopause velocity. An analytical error estimate for the normal velocity, which then folds into the thickness estimate, can be derived from equation (8.24) in *Sonnerup and Scheible* [1998]. This is routinely calculated in our procedure. Cases with large errors are excluded by our quality criteria (see section 4).

For each crossing, we manually inspected the magnetic profile and determined the jump in the maximum variance component of the magnetic field, $\Delta B_L$, and the duration, $\Delta T$, defined by a 76% change in the magnetic field as illustrated in Figure 3d. The magnetopause thickness, $D$, is given by $D = V_n \Delta T$. The motivation for using 76% of the full magnetic field jump is based on the assumption that the change in the maximum variance field component at the magnetopause, at least for high-magnetic shear, can be represented by a one-dimensional hyperbolic tangent profile, similar to the one-dimensional Harris-type current sheet [*Harris*, 1962].

In order to facilitate comparison with other regions and other studies, it can be useful to normalize the thickness to the local gyroradius ($r_g = T_n/|B|$). Both B field and temperature vary across magnetopause, so as a representative local gyroradius we first calculated the gyroradius at each individual sample within the 2 min interval around the crossing, and thereafter calculated the average of this. Perpendicular plasma temperature is only available from Cluster C1 and C3, so this normalization is only possible for around half of all crossing observations. Absolute gyroradii vary from around 30 km up to almost 900 km in our data set, with slightly larger average gyroradii at dawn (mean $r_{g_{DUSK}} = 111$ km, mean $r_{g_{DAWN}} = 116$ km).

As a supplement to the curlometer method, we can also calculate the current density, $J$ from a simple, one-dimensional version of Ampères law :

$$J = \frac{\Delta B_L}{\mu_0 D} = \frac{\Delta B_L}{\mu_0(V_n \Delta T)} \qquad (3)$$

With this method we can also estimate the current density for events where the curlometer method is not applicable (i.e. if only 1, 2, or 3 spacecraft crosses the magnetopause) or, alternatively, verify curlometer results. Applied to the measurements shown in Figure 2, this simplified method gives current densities ranging from 42 to 48 nA m$^{-2}$ for the four spacecraft, i.e., lower than the peak current but close to the calculated average curlometer current.

Some readers may ask why we did not utilize the four-spacecraft timing method, in which the time of a common feature (typically a sharp gradient in a measured quantity) and the spacecraft position are used to determine the orientation and normal speed of the magnetopause from triangulation. The answer to this is twofold.

First, the timing method is limited by many of the same constraints as the curlometer method described above. In addition to an appropriate spacecraft configuration, all four spacecraft need to cross the boundary, and the boundary needs to be stationary and planar (although, e.g., *Blagau et al.*, 2010 have outlined a method to address nonplanarity using four-spacecraft methods).

Second, during preparation of a companion paper (Anekallu et al., in preparation), two of us investigated a number of crossings independently. As this was a "sanity check," we deliberately agreed to agree on a specific method to determine the precise timing of the magnetopause crossings (though we both ended up using high-resolution magnetic field time profiles for timing). For cases with a well-defined crossing profile, the calculated quantities agreed well between us, but for more complicated profiles, we often ended up using different timing and thus obtained different boundary normals and magnetopause velocity. Another interesting outcome of the Anekallu et al., study is that when timing methods work well, the orientation is often very similar to that of the MVABC method.

### 3.3. Classification of Discontinuity Type

In a fluid description, the magnetopause can be described either as a tangential discontinuity (TD) or as a rotational discontinuity (RD) [see e.g., *Hudson*, 1970]. A TD implies a complete separation of two plasma regimes (in this case the magnetosheath on one side and the magnetosphere on the other side). The boundary as a whole may move, but there is no transport of plasma across the discontinuity, and there is no magnetic field along the boundary normal. An RD-like magnetopause, on the other hand, implies transport across the boundary and a normal magnetic field, and indicates the presence of reconnection. In the vicinity of the reconnection site (magnetic X line), the plasma flow is Alfvénic, i.e., the Walén relation [see e.g., *Sonnerup et al.*, 1987; *Paschmann and Sonnerup*, 2008] is satisfied:

$$\vec{v} - \vec{V_{HT}} = \pm \vec{V_A} \tag{4}$$

where $\vec{v}$ is the plasma flow, $\vec{V_{HT}}$ is the deHoffmann-Teller frame velocity discussed above (section 3.2). $\vec{V_A}$ is the local Alfvén velocity.

A simple check of the Walén test can be done by plotting (component by component) the plasma flow in the HT frame against the Alfvén velocity. Provided that the magnetopause is indeed planar and that the deHoffmann-Teller frame is properly defined, a linear regression line with slope near $\pm 1$, suggests an RD-like magnetopause if one is able to estimate the Alfvén speed properly. Correspondingly, a regression line with a slope close to 0 indicates a TD-like magnetopause.

For an RD, the flow across the boundary is proportional to the normal magnetic field, i.e., $v_n \propto B_n$. A positive (negative) slope of the regression means that normal magnetic field and flow have the same (opposite) signs. At the magnetopause, we can assume that the flow is inward, i.e., from the magnetosheath into the magnetosphere. Our use of constrained minimum variance analysis implies that $B_n$ is zero by definition, but the sign of the Walén slope tells us the sign of the normal magnetic field and provides information about the magnetic topology around the X line. A positive Walén slope indicates a crossing sunward of the X line, whereas a negative slope indicates a crossing on the tailward side [*Paschmann et al.*, 2005].

## 4. Characteristics of the Magnetopause Flanks

In this section, we make use of the large data set to characterize some of the key macroscopic parameters of the magnetopause. We also check the correlation with these distributions in order to determine how the IMF and solar wind affect the flank magnetopause.

In the presentation of the individual results we do not show error bounds or statistical spread (in section 5, we will provide statistical spreads in the overall moments, though). Error estimates exist for both single-spacecraft and multispacecraft methods, but such estimates typically only account for statistical errors in the methodology, and typically depend strongly on the number of samples entering. Errors due to breakdown of the underlying assumption are not taken into account but are typically dominating though difficult to quantify [e.g., *Zhou et al.*, 2009; *Vogt et al.*, 2011].

For classification of discontinuity type and to express the thickness in terms of ion gyroradii, we make use of plasma density and perpendicular temperature from the CIS moments. The presence of cold plasma (i.e., ions with energies below the effective instrument energy threshold) and heavy ions will affect the accuracy of these moments. Section 5.3 of the preceding *Haaland and Gjerloev* [2013] paper contains a discussion about error sources and confidence of the results from that study. The same argumentation also applies to the extended data set used in the present study.

Note that when we investigate the role of the solar wind and IMF in the sections below, we use the prevailing *upstream* solar wind as extracted from NASA/Goddard Space Flight Center's OMNI data set [*King and Papitashvili*, 2005] through Coordinated Data Analysis Web (CDAWeb) (http://omniweb.gsfc.nasa.gov/index.html). Neither propagation time from the upstream position to Cluster nor possible time history effects of the solar wind have been taken into account, as both these factors are difficult to quantify.

In addition to visual inspection of each event, we also applied a set of formal quality criteria to ensure that the calculation of orientation, velocity, thickness, and current density were as accurate as possible. Records with obvious erroneous values, typically caused by breakdown of model assumptions or data gaps, were discarded.

Table 1 lists the filter criteria we used to select data for the characterization. The values are largely adapted from *Haaland and Gjerloev* [2013], but we have adjusted the requirement for deHoffmann-Teller correlation coefficient to facilitate comparison with the studies by *Paschmann et al.* [2005] and *Chou and Hau* [2012]. Small adjustments in any of the above filter criteria had no significant impact on the results. The quality criteria are quite conservative, and after having filtered the data according to these criteria, we ended up with a substantially smaller number of suitable crossings. But the remaining events provide us with a subset of crossings where reliable estimates of the key magnetopause parameters were possible.

### 4.1. Magnetopause Thickness

Figure 4a shows a histogram of magnetopause current sheet thicknesses for dawn (red color) and dusk (blue color). For convenience, we also show the accumulated distribution as lines, with the scale given along the secondary vertical axis at right. The number of crossings on each flank and the mean and median thickness for each flank are given in the plot legend. Due to the orbit of Cluster and the aberration of the magnetotail mentioned in section 2, we have a larger numbers of crossings at dawn.

From the data, we obtain an overall mean (median) magnetopause thickness of 1629 (1289) km. The magnetopause is thicker at dawn than at dusk: 1699 km at dawn and 1496 km at dusk. Median thicknesses are 1396 and 1145 km at dawn and dusk, respectively.

For comparison *Russell and Elphic* [1978], using dayside observation from the ISEE-1 and ISEE-2 satellites reported thicknesses in the range 500–1000 km. A subsequent study by *Berchem and Russell* [1982] used an even larger data set and obtained an average thickness of the dayside magnetopause of 923 km. Note that these studies used the full 100% transition (see definition in section 3) of the magnetopause current sheet in their estimation, though. *Phan and Paschmann* [1996], using Active Magnetospheric Particle Trace Explorers/Ion Release Module (AMPTE/IRM) measurements, also reported an average thicknesses around 900 km for the dayside magnetopause.

Our results clearly indicate that the flank magnetopause is thicker than the dayside magnetopause. This is in contrast to the findings by *Paschmann et al.* [2005]. They reported an average thickness of $\simeq 750$ km and attributed the lack of systematic growth of thickness with distance from the subsolar point to a low-current diffusion coefficient. Note, however, that the *Paschmann et al.* [2005] study was based on a single day of observations from July 2001, whereas the present study is based on 10 years of data. If we limit our data set to the same interval as their study, we obtain similar results.

To study the effect of solar wind parameters on magnetopause thickness, we plotted magnetopause thickness versus several solar wind parameters. In Figure 4b, we show magnetopause thickness versus the solar wind dynamic pressure. There seems to be little or no correlation between thickness and solar wind pressure. Correlating with solar wind density or solar wind bulk velocity (not shown) does not reveal any systematic dependences either. Nor do we find any strong correlations between magnetopause thickness and IMF orientation.

Noting that the dawn and dusk boundary conditions, e.g., plasma temperature and magnetic field strength can be different [e.g., *Walsh et al.*, 2013, and references therein], we also tried to normalize the magnetopause thickness to the local ion gyroradius. Figure 4c shows a histogram of the thickness expressed in terms of number of local gyroradii. The majority of crossings are 5–15 ion gyroradii thick. Only a very few crossings have thicknesses less than 2 ion gyroradii, but since our methodology and quality settings are based on an MHD approach, this is hardly surprising. There are also some crossings with thicknesses $\geq 30\,r_g$, so the lines indicating cumulative values do not reach 100% within the axis range shown.

A significant dawn-dusk asymmetry is still apparent after normalization to the local gyroradii. The mean dawn magnetopause is 19 gyroradii thick, whereas the mean dusk magnetopause is 15 gyroradii thick. Corresponding median thicknesses are 13 and 10 gyroradii for dawn and dusk, respectively. This result indicates that other factors, such as intrinsic properties of the current sheet, are at least partially responsible for the difference in thickness between the dawn and dusk magnetopause.

### 4.2. Magnetopause Motion and Location

The magnetopause is known to move rapidly in and out. This is partly due to dynamical changes in the solar wind but is also observed locally as a consequence of surface waves. To characterize this motion, Figure 5 shows histograms and solar wind dependencies of the magnetopause velocity (with velocity, we here mean the motion of the magnetopause along its normal direction).

Mean and median magnetopause normal velocities are 56 and 41 km s$^{-1}$, respectively. Our data set contains crossings where the magnetopause moved with more than 100 km s$^{-1}$, but the majority of the events moved with velocities below 40 km s$^{-1}$. Once again, we see a dawn-dusk asymmetry. Normal velocities at dawn are higher than at dusk. Mean and median dawn velocities are 64 km s$^{-1}$ and 48 km s$^{-1}$, respectively. At dusk, the corresponding values are 42 km s$^{-1}$ and 32 km s$^{-1}$, respectively.

The distribution in Figure 5 resembles the results by *Paschmann et al.* [2005], who reported a mean velocity of 48 km s$^{-1}$ for their 96 dawn magnetopause crossings. As in their study, we find rather few events with very fast ($\geq$ 100 km s$^{-1}$) or very slow ($\leq$ 10 km s$^{-1}$) motion. By contrast, *Plaschke et al.* [2009], using observations from the five Time History of Events and Macroscale Interactions during Substorms spacecraft, mostly on the dayside, found a number magnetopause crossings with velocities less than 10 km s$^{-1}$. Differences in methodology may explain parts of the discrepancy between these studies: *Plaschke et al.* [2009] used an indirect method based on magnetic field observations and timing only, whereas *Paschmann et al.* [2005] used a modified multispacecraft technique.

Figure 5b shows the magnetopause motion as function of solar wind ram pressure. We also checked the correlations between velocity and solar wind velocity and density, but these are not shown and show no significant correlations. As for the thickness study above, there seems to be only a weak statistical correlation between the magnetopause velocity at the flanks and prevailing upstream solar wind conditions. Except for a slightly higher average velocity at dawn for positive IMF *By* conditions, we do not find any significant correlation between magnetopause motion and IMF orientation either.

Given that the geomagnetic field is fairly rigid, one would expect that the position and motion of the magnetopause were mainly controlled by the impact of the solar wind, in particular the dynamic pressure. Erosion, and thus to some extent the thickness of the magnetopause, is believed to be a result of reconnection, primarily governed by the magnitude and orientation of the IMF. The dynamic pressure probably affects the magnetopause thickness as well (see discussion in section 4.5), but due to its low total mass, changes in the dynamic pressure are more likely to compress it [*Sonnerup et al.*, 2006]. The latter is corroborated by Figure 1 where we plotted the positions of low-latitude (within ±10° of the ecliptic plane) C3 observations of magnetopause encounters where the filter criteria in Table 1 were satisfied. Red dots indicate crossings during times when the solar wind dynamic pressure was above average (2.23 nPa for this subset); black dots indicate corresponding crossing positions during periods with lower than average solar wind dynamic pressure. The magnetopause is observed closer to Earth both at dawn and dusk during periods with high-solar wind pressure.

From Figure 1, it is also apparent that the magnetopause position is very variable and can move significantly as response to solar wind variations. Some of the crossings were observed more than 5 *Re* further inward than the magnetopause position predicted by the *Fairfield* [1971] model. Due to Cluster's ≃20 *Re* apogee, we probably miss a number of magnetopause traversals tailward of the terminator, in particular on dusk. Our data set is therefore not very well suited for testing and benchmarking of modern magnetopause models such as those of *Lin et al.* [2010] that discuss asymmetries. A comprehensive test of magnetopause models using Cluster was reported by *Case and Wild* [2013], although they only used dayside crossings.

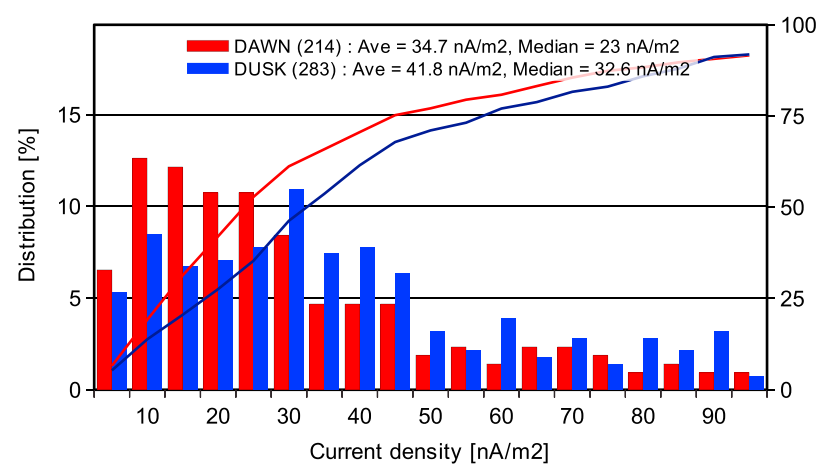

**Figure 6.** Distribution of magnetopause current densities based on the curlometer results. Each bin in the histograms are 5 nA m⁻² wide. Lines show cumulated values and refer to the secondary vertical axis at right. Dawn crossings are characterized by lower current densities than their dusk counterparts.

### 4.3. Current Density

Current density calculations using the curlometer method is possible for the years 2001–2006. After that, the Cluster interspacecraft separation was changed, and the tetrahedron-like configuration required to calculate gradients were rarely achieved. Magnetopause current densities and their relation to other current systems and response to geoactivity were already extensively discussed in *Haaland and Gjerloev* [2013], so here we just briefly repeat some of the results from that paper.

Figure 6 shows a histogram of current densities similar to those in *Haaland and Gjerloev* [2013], but this time without any sorting according to geomagnetic activity. Each bin in the histograms is 5 nA m⁻² wide. Red bars indicate dawn current densities, and blue bars indicate dusk current densities. From the distribution in Figure 6, it is clear that dusk crossings are characterized by higher current densities than their dawn counterparts. Mean current densities are 42 and 35 nA m⁻² for dusk and dawn, respectively. Median current densities are 33 and 23 nA m⁻².

As pointed out in *Haaland and Gjerloev* [2013], the dawn-dusk asymmetry tended to be larger during disturbed conditions (see their Figure 7). As possible explanations for the dawn-dusk asymmetry and its dependence on geoactivity, *Haaland and Gjerloev* [2013] discussed two effects. First, influences from the ring current would cause a larger difference between the internal (i.e., magnetospheric) and external (magnetosheath) magnetic field on dusk. This effect would be more pronounced during disturbed conditions when the ring current is stronger and has a stronger asymmetry [*Newell and Gjerloev*, 2012]. A second possible explanation for the observed current density asymmetry is that the external conditions are different at dawn and dusk. In particular, the magnetic field and fluctuations in the magnetic fields have been found to be larger at dusk [*Fairfield and Ness*, 1970; *Shevyrev et al.*, 2007; *Walsh et al.*, 2012].

For a number of magnetopause crossings, the quality criteria in Table 1 were fulfilled for both the curlometer method (equation 1) and the single-spacecraft current determination (equation 3), so that a consistency check was possible. The dawn-dusk asymmetry is also apparent from the single-spacecraft methods. Current densities calculated with the curlometer method were typically around 40% higher than those obtained from the single-spacecraft method. However, note that the curlometer values used in our characteristics are the peak currents (see Figure 2d), whereas the single-spacecraft method returns an average over the current sheet thickness.

### 4.4. Tangential or Rotational Discontinuity

In our analysis, we routinely recorded the HT frame quality and the Walén slope. For about 1700 cases, we considered the frame quality to be sufficient (HT correlation coefficient ≥ 0.85—same threshold as used by *Paschmann et al.* [2005]) enough to trust the calculated Walén regression line slopes. Figure 7a shows a histogram of the slope values. Note that the Walén test requires plasma density to calculate the Alfvén velocity. The density was not available for all crossings, so the MHD classification is based on fewer events than the thickness and motion calculations.

Recall that Walén slopes close to 0 indicate a tangential discontinuity whereas values close to ± 1 indicate a rotational discontinuity. For various reasons, including pressure anisotropy and composition effects (see discussion in, e.g., *Paschmann and Sonnerup* [2008] about this), values close to unity are rarely observed, however. Both *Paschmann et al.* [2005] and *Chou and Hau* [2012] therefore classified cases with Walén slopes greater than 0.5 as RDs. We will adapt these thresholds also for classifications of our events.

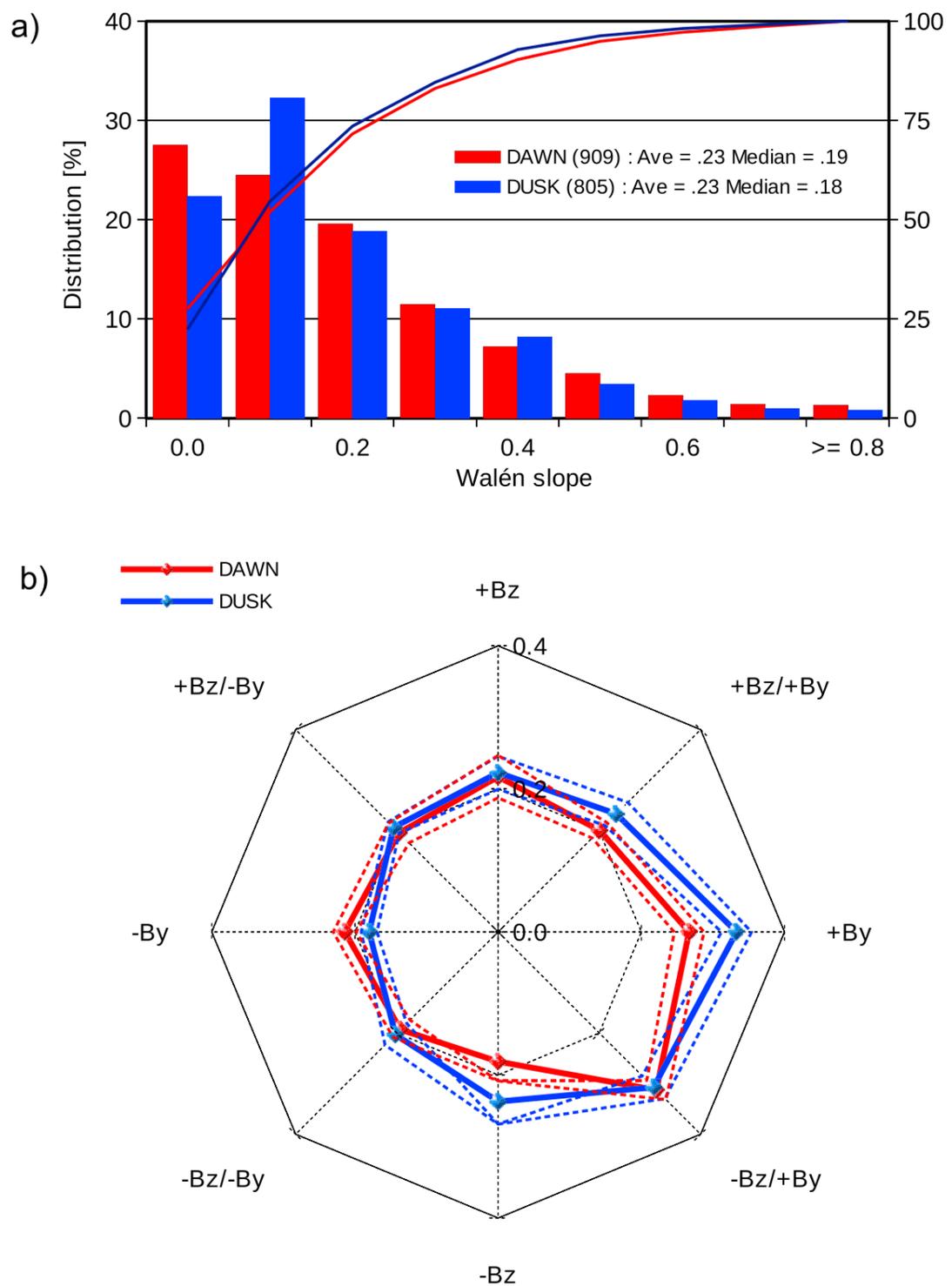

**Figure 7.** (a) Histograms of Walén regression line slopes. Values close to zero indicate a tangential discontinuity, and values close ± 1 indicate a rotational discontinuity. (b) Solid lines indicate average Walén slopes as function of IMF direction. Dashed lines indicate the statistical spread (standard error, defined as $\sigma / \sqrt{N}$, where $\sigma$ is the standard deviation and $N$ is the number of observations).

There is no significant difference between the dawn and dusk flank when it comes to classification. From Figure 7a it is clear that the large majority are TD-like boundaries at both flanks. Only 124 cases, corresponding to $\simeq 7\%$ of the total had a slope greater than 0.5. If we take a more conservative approach and define crossings with Walén regression line slopes below 0.2 as TDs and only crossings with slopes $\geq 0.8$ as RDs, we obtain 52% TDs and only about 1.3% RDs.

For comparison, the above cited *Paschmann et al.* [2005] study was able to classify 60 of their dawn flank magnetopause crossings and concluded that 34% were RDs. *Chou and Hau* [2012], using 384 predominantly dayside magnetopause crossings from the AMPTE/IRM spacecraft, reported 8–12% RDs, depending on selection criteria.

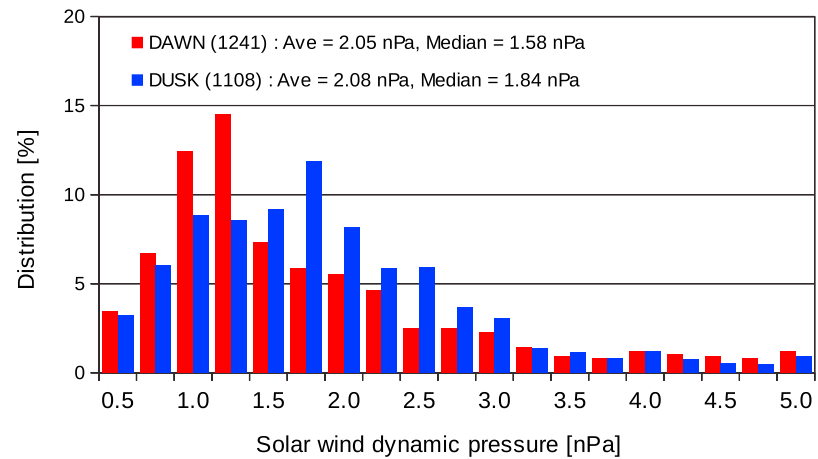

**Figure 8.** Distribution of solar wind dynamic pressure. Each bin in the histograms is 0.25 nPa wide. The mean solar wind dynamic pressure is almost the same for dawn and dusk crossings. The corresponding median moments are slightly more asymmetric, but the solar wind dynamic pressure alone is not enough to explain the observed dawn-dusk differences in magnetopause parameters.

Figure 7b shows how the average Walén slopes depend on IMF direction. Each branch of this spider-like plot represents a 45° IMF clock angle sector (where the IMF clock angle is defined as $\tan^{-1}(B_{Z\,IMF}/B_{Y\,IMF})$) around the IMF orientations indicated in the plot. For the dayside magnetopause, one would expect enhanced reconnection and Walén slopes closer to unity for southward IMF. Our results in Figure 7b suggest a slightly higher values for the combination negative IMF $Bz$/positive IMF $By$, but average slope values are still small. We also checked the IMF dependence of the other parameters (thickness, current density, and velocity) and obtained similar, nonconclusive results. It thus seems that the prevailing IMF orientation has less impact on properties at the flank magnetopause than at the dayside.

As mentioned in section 3.3, a positive sign of the Walén slope indicates a crossing sunward of an $X$ line, whereas a negative slope indicates a crossing on the tailward side [*Paschmann et al.*, 2005]. In our data set, most of the slopes have a negative sign, although for absolute slope values above 0.5, there is roughly equally many cases with positive sign as with negative sign.

A majority of negative slopes would indicate that any reconnection activity is more likely to take place sunward of the spacecraft. This, again, would indicate that the IMF interaction with the geomagnetic field on the dayside is the main cause for reconnection, rather than rolled-over KH waves, which we would expect to present further downtail along the flank, and in particular on the duskside [*Taylor et al.*, 2012].

### 4.5. The Role of the Solar Wind and IMF

As noted in section 4.2, position, motion, and current density of the magnetopause, at least on the dayside, are largely controlled by the impact of the solar wind. It is therefore natural to examine whether any of the dawn-dusk asymmetries discussed in the previous sections can be explained by corresponding asymmetries in the solar wind or a bias in our collection of events.

Figure 8 shows the distribution of the solar wind dynamic pressure. In our 10 year collection of magnetopause observations, the mean solar wind dynamic pressure is slightly higher for the dusk crossings (mean solar wind dynamic pressure = 2.08 nPa versus 2.05 nPa at dawn; corresponding median values are 1.84 and 1.58 nPa, respectively). We argue that this is not sufficient to explain the dawn-dusk asymmetries in the observed magnetopause properties.

For the magnetopause thickness, $D$, an adiabatic compression, $\Delta D$, due to solar wind pressure change, $\Delta P_{DYN}$, would be of the order of

$$\frac{\Delta D}{D} = (-3/5)\frac{\Delta P_{DYN}}{P_{DYN}} \qquad (5)$$

A 15% difference in solar wind pressure between dawn and dusk would only be able to account for maximum 10% difference in thickness. Magnetic forces inside the magnetopause current layer further diminish the effect of the solar wind pressure. The same argument also applies to current density.

Likewise, magnetopause *motion* is primarily controlled by sudden changes, $\delta P_{DYN}$, in the solar wind dynamic pressure. For a magnetopause with thickness $D$ and plasma mass density $\rho$, the acceleration of the magnetopause, d$v$/d$t$, would be of the order of

$$\frac{dv}{dt} = \frac{\delta P_{DYN}}{\rho D} \qquad (6)$$

**Table 2.** Summary of Statistical Moments[a]

| Parameter | Method | Both | | | | Dawn | | | | Dusk | | | |
|---|---|---|---|---|---|---|---|---|---|---|---|---|---|
| | | Mean | Median | SErr | (N) | Mean | Median | SErr | (N) | Mean | Median | SErr | (N) |
| Thickness (km) | Single | 1628 | 1289 | 30.2 | (1756) | 1699 | 1396 | 37.8 | (1073) | 1496 | 1145 | 49.8 | (583) |
| Thickness ($r_g$) | Single | 17.1 | 11.9 | - | (711) | 18.9 | 13.1 | - | (408) | 14.6 | 9.9 | - | (303) |
| Velocity (km s$^{-1}$) | Single | 56.3 | 41.1 | 1.2 | (1756) | 64.4 | 47.9 | 1.6 | (1073) | 41.5 | 31.6 | 1.5 | (583) |
| $J_{MAX}$ (nAm$^{-2}$) | Multi | 38.7 | 28.6 | 1.5 | (497) | 34.7 | 23.0 | 2.2 | (214) | 41.8 | 32.6 | 1.9 | (283) |

[a]The thickness is given both in units of kilometer and as the ratio between the thickness and the local ion gyroradius. The SErr column shows the standard error of the distribution used to calculate the moments. It is defined as $\sigma/\sqrt{N}$, where $\sigma$ is the standard deviation and $N$ is the number of observations, listed in brackets.

The 1 min OMNI data set used as solar wind parameters does not really contain any good proxies for $\delta P_{DYN}$, but we have no reason to believe that our data set is biased toward higher $\delta P_{DYN}$, for one of the flanks.

These estimates suggest that the solar wind dynamic pressure or the dynamics of the solar wind alone is not enough to explain the observed dawn-dusk differences in magnetopause parameters. This is also corroborated by Figures 4b and 5b which show that neither thickness nor velocity depend strongly on the solar wind dynamic pressure.

The IMF clock angle distribution is dominated by $\pm$ IMF *By* orientations as expected from the Parker spiral. Our dusk collection contains more events during negative IMF *By* values, whereas the dawn collection contains more events during positive IMF *By*. We find no strong correlations between IMF orientation and any of the investigated macroscopic magnetopause properties, however. The thicker magnetosheath layers at the flanks seem to attenuate and filter out much of the direct IMF and solar wind impact.

## 5. Summary and Conclusion

We have used an extensive number of Cluster traversals of the flank magnetopauses to characterize macroscopic features of the dusk and dawn magnetopause. Using a combination of single-spacecraft and multispacecraft methods, we were able to calculate velocity, orientation, and thickness of the magnetopause for a large number of magnetopause crossings during the years 2001 to 2010. For the years 2001 to 2006, we were also able to use the four-spacecraft curlometer method to derive detailed current profiles for a number of crossings. Table 2 summarizes the moments of the calculations. In addition to the full averages, we also show the dawn and dusk moments separately.

An interesting result of the study is the persistent dawn-dusk asymmetry observed. The dusk magnetopause is typically thinner and has a higher-current density than its dawn counterpart. Conversely, the typical dawn magnetopause is thicker, and with a lower current density. Consequently, the total current carried is similar for the two flanks. The asymmetry remains if we normalize the thickness to the local gyroradius. This pronounced dawn-dusk asymmetry was pointed out by *Haaland and Gjerloev* [2013], but to our knowledge, no complete explanation for this asymmetry exists. The solar wind dynamic pressure or the orientation of the IMF cannot alone explain the observed asymmetries. Other external factors, in the form of different plasma properties in the dusk and dawn magnetosheath most likely play a significant role. Properties and processes inside the magnetosphere, e.g., asymmetries in the ring current and the coupling to the ionosphere, probably also contribute to the observed magnetopause dawn-dusk asymmetry. And, finally, an intrinsic asymmetric dependence of magnetopause structure on the sense of the flow relative to the magnetic field may play a role.

The magnetopause velocity also shows large variations. Typical magnetopause velocities are in the range 10–40 km s$^{-1}$, but velocities in excess of 200 km s$^{-1}$ were observed for some of the crossings. In contrast to the dayside magnetopause study by *Plaschke et al.* [2009], we find few magnetopauses moving with velocities below 10 km s$^{-1}$, though differences in methodology and selection may play a role. The dawnside magnetopause was on average moving faster than on dusk, which may indicate that the dawn flank is more turbulent, possibly due to asymmetries in bow shock geometry between dawn and dusk [*Walsh et al.*, 2013].

Current sheet profiles obtained from the curlometer method demonstrate that the magnetopause cannot in general be considered as a single current sheet. Rather, many of the crossings indicated layered structures consisting of two or more adjacent current sheets, often with different current orientation. This multilayer structure is seen in typical kinetic models of the tangential discontinuity magnetopause [e.g., *Roth et al.*, 1996]. Typical ion gyroradii are of the order 80–100 km in this region, so for all events, the magnetopause current sheet thickness is several ion gyroradii thick.

Using the same classification scheme as in *Paschmann et al.* [2005] and *Chou and Hau* [2012] we also tried to classify the nature of the magnetopause current sheet. The large majority of crossings seems to be of a tangential discontinuity type, with little or no transport across the boundary. Only about 7% of the cases had a Walén slope indicating a rotational discontinuity.

**Acknowledgments**

Computer code used for the calculations in this paper has been made available as part of the QSAS science analysis system. QSAS is provided by the United Kingdom Cluster Science Centre (Imperial College London and Queen Mary, University of London) supported by the United Kingdom Science and Technology Facilities Council (STFC). Solar wind data were obtained from the Coordinated Data Analysis Web (CDAWeb—see http://cdaweb.gsfc.nasa.gov/about.html). We also thank the International Space Science Institute, Bern, Switzerland for providing computer resources and infrastructure for data exchange.

Yuming Wang thanks the reviewers for their assistance in evaluating this paper.